\documentclass[preprintnumbers,prd,showpacs,floatfix,superscriptaddress,nofootinbib, letterpaper]{revtex4-2}

\usepackage{graphicx} 
\usepackage{xcolor}
\usepackage{amsmath, amssymb}
 \usepackage{hyperref}
\usepackage[top = 2cm, bottom = 2cm, right = 2cm, left =2cm]{geometry}
\usepackage[utf8]{inputenc}
\usepackage{textgreek}
\usepackage{float}

\DeclareUnicodeCharacter{202F}{\,}
\begin{document}

\title{Geometrically Regular Black Object Solutions in Lower-Dimensional Gauss–Bonnet Gravity and Its Unimodular Extension}

\author{G. Alencar}
\email{geova@fisica.ufc.br} 
\affiliation{Departamento de F\'isica, Universidade Federal do Cear\'a, Caixa Postal 6030, Campus do Pici, 60455-760 Fortaleza, Cear\'a, Brazil}

\author{T. M. Crispim} 
\email{tiago.crispim@fisica.ufc.br}
\affiliation{Departamento de F\'isica, Universidade Federal do Cear\'a, Caixa Postal 6030, Campus do Pici, 60455-760 Fortaleza, Cear\'a, Brazil}

\author{C. R. Muniz}
\email{celio.muniz@uece.br}
\affiliation{Universidade Estadual do Cear\'a, Faculdade de Educa\c c\~ao, Ci\^encias e Letras de Iguatu, 63500-000, Iguatu, CE, Brazil.}

\date{\today}

\begin{abstract}
We investigate the construction of regular compact objects in the recently proposed lower-dimensional Einstein--Gauss--Bonnet (EGB) gravity obtained through regularized dimensional reduction. Unlike the standard BTZ black hole, the corresponding vacuum EGB solution develops a genuine curvature singularity at the origin, providing an interesting setting in which higher-curvature corrections deteriorate the ultraviolet behavior of spacetime. To address this issue, we reconstruct matter sectors capable of restoring regularity while preserving the BTZ-like asymptotic structure. First, we derive regular black-hole solutions supported by nonlinear electrodynamics and determine the corresponding electromagnetic Lagrangians directly from the field equations. We then extend the analysis to Simpson--Visser black-bounce geometries, obtaining smooth throat configurations with finite curvature invariants throughout the spacetime. As an alternative regularization mechanism, we formulate a unimodular extension of lower-dimensional EGB gravity and show that standard Maxwell fields can support regular geometries through a dynamical exchange between the vacuum and matter sectors mediated by a spacetime-dependent cosmological function. We further investigate the thermodynamic properties of the regular black-hole and black-bounce solutions, showing that the matter sector modifies the evaporation process, allows for remnant formation, and produces nontrivial phase transitions. In the black-bounce case, the thermodynamic quantities smoothly recover the EGB-BTZ behavior in the appropriate limit. These results demonstrate that lower-dimensional EGB gravity provides a useful laboratory for exploring the interplay between higher-curvature corrections, regular compact objects, nonlinear electrodynamics, and unimodular gravity.

\end{abstract}

\maketitle

\section{Introduction}

The search for a consistent framework that generalizes Einstein's general relativity while maintaining second-order field equations leads naturally to Lovelock gravity \cite{Lovelock:1971yv}. In this theory, the gravitational action is extended by a series of higher-order curvature invariants, where the first non-trivial addition is the Gauss--Bonnet (GB) term, defined as:
\begin{equation}
\mathcal{G} = R_{abcd}R^{abcd} - 4R_{ab}R^{ab} + R^2.
\label{GBinv}
\end{equation}
Although the GB invariant plays a crucial role in higher-dimensional physics, it behaves as a total derivative in four dimensions, meaning it does not influence the classical dynamics in $D=4$. This limitation was recently challenged by a proposal to rescale the coupling constant as $\alpha \to \alpha/(D-4)$, an approach intended to capture non-trivial effects in the limit $D \to 4$ \cite{Glavan:2019inb}. While this sparked a wave of new solutions for black holes and cosmology \cite{Glavan:2019inb,Kumar:2020uyz,Fernandes:2020rpa,Cunha:2025jzh}, it also faced criticism regarding its mathematical consistency and the lack of a fully covariant description in the limit \cite{Gurses:2020ofy,Hennigar:2020lsl,Arrechea:2020evj,Fernandes:2022zrq}.

To resolve these ambiguities, regularized dimensional reduction schemes were developed, establishing that the consistent $D \to 4$ limit of Einstein--Gauss--Bonnet (EGB) gravity belongs to the Horndeski class of scalar--tensor theories \cite{Lu:2020iav,Mann:1992ar,Fernandes:2020nbq}. The resulting action is expressed as:
\begin{equation}
\label{SD}
S = \int d^D x \sqrt{-g} \Bigl[ R - 2\Lambda + \alpha \Bigl( \psi\mathcal{G} + 4G^{ab}\partial_a\psi\partial_b\psi - 4(\partial\psi)^2\Box\psi + 2((\nabla\psi)^2)^2 \Bigr) + \mathcal{L}_M \Bigr].
\end{equation}

This formulation provides a reliable foundation not only for $D=4$ but also for lower-dimensional gravity, which serves as a vital theoretical laboratory. A prime example is the BTZ black hole \cite{Banados:1992wn}, a fundamental object for exploring holography and the quantum nature of spacetime \cite{Carlip:1995qv,Konoplya:2004ik}. Recently, a vacuum BTZ-like solution within this regularized EGB framework was derived in \cite{Hennigar:2020fkv}:
\begin{equation}
f(r) = -\frac{r^2}{2\alpha} \left[ 1 - \sqrt{ 1 + \frac{4\alpha}{r^2} \left( \frac{r^2}{\ell^2} - m \right) } \right],
\label{NewSolution}
\end{equation}
which converges to the standard BTZ geometry when $\alpha \to 0$. 

The appearance of curvature singularities remains one of the most persistent challenges in gravitational physics. Since the pioneering proposal of regular black holes by Bardeen, considerable effort has been devoted to constructing geometries in which curvature invariants remain finite throughout the spacetime while preserving the existence of event horizons. Various mechanisms have been proposed to achieve this goal, including nonlinear electrodynamics, modified gravity theories, quantum-inspired corrections, and effective matter sectors \cite{Ayon-Beato:1998hmi,Ayon-Beato:2000mjt,Hayward:2005gi,Maluf:2022jjc,Muniz:2025ugk,Pinto:2025loq}. Besides regular black holes, growing attention has recently been devoted to black-bounce geometries, which interpolate continuously between black holes and traversable wormholes and provide a particularly useful framework for exploring singularity resolution beyond classical general relativity \cite{Simpson:2018tsi,Franzin:2021vnj,Muniz:2024wiv,Lobo:2020ffi,Furtado:2022tnb}.

However, unlike the standard BTZ black hole, which is locally AdS and free of curvature singularities, the vacuum solution \eqref{NewSolution} is geometrically singular at the origin, as explicitly shown in \cite{Hennigar:2020fkv}. In particular, the Kretschmann scalar diverges in the limit $r\to0$, indicating that the Gauss--Bonnet corrections generate a genuine curvature singularity even in the absence of matter fields. This result is especially intriguing because the original BTZ geometry is regular everywhere, implying that the higher-curvature sector effectively deteriorates the ultraviolet behavior of spacetime. Such a feature naturally raises the question of whether physically acceptable matter sources can counterbalance these higher-curvature effects and restore the regularity properties of the BTZ geometry while preserving its asymptotic structure. Addressing this problem may provide new insights into the interplay between higher-curvature corrections and regular compact objects in lower-dimensional gravity. Motivated by these considerations, in the present work we construct explicit regular black hole and black-bounce solutions supported by nonlinear electrodynamics sectors and investigate their geometric and thermodynamic properties.

While curvature singularities are traditionally bypassed through nonlinear electrodynamics (NED), recent developments suggest that alternative mechanisms may also support regular compact objects. In particular, unimodular gravity has attracted renewed interest as a framework in which the cosmological sector emerges dynamically through a restricted variation of the metric determinant \cite{Ellis:2010uc,Alencar:2026hjf,Alencar:2026oxy}. In this approach, the effective cosmological function may exchange energy with the matter sector, relaxing the standard conservation law and opening new possibilities for the construction of regular geometries. Remarkably, recent studies have shown that this mechanism can sustain regular black holes sourced solely by standard Maxwell fields, without invoking nonlinear electromagnetic interactions \cite{Alencar:2026hjf,Alencar:2026oxy}. Motivated by these results, we investigate whether a similar regularization mechanism can be successfully implemented in the context of lower-dimensional Einstein--Gauss--Bonnet gravity.

Our manuscript is organized as follows. Section II discusses the reconstruction of the NED sector required for regular BTZ black holes. In Section III, we extend our analysis to black bounce geometries of the Simpson--Visser type. Section IV introduces the unimodular extension of lower-dimensional EGB gravity and demonstrates how regular solutions emerge from a Maxwell field. Thermodynamic properties  are evaluated in Section V. Finally, Section VI summarizes our conclusions.

\section{Regular Black Holes Supported by Nonlinear Electrodynamics}
In this section, we will consider the action (\ref{SD}) and spherical symmetry, with metric 
\begin{equation}\label{bh3}
    ds^2=-f(r)dt^2+\frac{1}{f(r)}dr^2+r^2d\varphi^2.
\end{equation}
The gravitational field equations are given by
\begin{equation}\label{eisteineq}
   G_{\mu\nu} + \Lambda g_{\mu\nu}-\alpha \mathcal{H}_{\mu\nu}=T_{\mu\nu}
\end{equation}
where $G_{\mu\nu}$ is the Einstein tensor 
and $T_{\mu}^{\;\nu}$ is the stress-energy tensor and $\mathcal{H}_{\mu\nu}$ is defined by \cite{Alencar:2026yyz}
\begin{eqnarray}
\mathcal{H}_{\mu\nu}\!&\!=\!&\!2R\left(\nabla_\mu\nabla_\nu \psi-\nabla_\mu\psi\nabla_\nu\psi\right)+2G_{\mu\nu}\left[(\nabla\psi)^2-2\Box\psi\right]+4G_{\nu\alpha}\left(\nabla^\alpha\nabla_\mu\psi-\nabla^\alpha\psi\nabla_\mu\psi\right)\nonumber\\
&&+4G_{\mu\alpha}\left(\nabla^\alpha\nabla_\nu\psi-\nabla^\alpha\psi\nabla_\nu\psi\right)
+4R_{\mu\alpha\nu\beta}\left(\nabla^\beta\nabla^\alpha\psi-\nabla^\alpha\psi\nabla^\beta\psi\right)+4\nabla_\alpha\nabla_\nu\psi\left(\nabla^\alpha\psi\nabla_\mu\psi-\nabla^\alpha\nabla_\mu\psi\right)\nonumber\\
&&+4\nabla_\alpha\nabla_\mu\psi\nabla^\alpha\psi\nabla_\nu\psi+4\Box\psi\nabla_\nu\nabla_\mu\psi
-4\nabla_\mu\psi\nabla_\nu\psi\left[(\nabla\psi)^2+\Box\psi\right]^2-g_{\mu\nu}\Big\{2R\left[\Box\psi-(\nabla\psi)^2\right]\nonumber\\
&&+4G^{\alpha\beta}\left(\nabla_\beta\nabla_\alpha\psi-\nabla_\alpha\psi\nabla_\beta\psi\right)+2(\Box\psi)^2-(\nabla\psi)^4+2\nabla_\beta\nabla_\alpha\psi\left(2\nabla^\alpha\psi\nabla^\beta\psi-\nabla^\beta\nabla^\alpha\psi\right)\Big\}.  
\end{eqnarray}
The field equation of the scalar field $\psi$, $\mathcal{E}_\psi = \delta S/\delta \psi = 0$ reads
\begin{eqnarray}
\label{r4EGBgfeq1}
\mathcal{E}_{\psi}=R^{\mu\nu}\nabla_{\mu}\psi\nabla_{\nu}\psi-G^{\mu\nu}\nabla_{\mu}\nabla_{\nu}\psi-\Box\psi\nabla_{\mu}\psi\nabla^{\mu}\psi+\nabla_{\mu}\nabla_{\nu}\psi\nabla^{\mu}\nabla^{\nu}\psi-(\Box\psi)^{2}-2\nabla_{\mu}\psi\nabla_{\nu}\psi\nabla^{\mu}\nabla^{\nu}\psi=0.    
\end{eqnarray} 

Since the laws of physics in a standard covariant formulation do not depend on the chosen coordinate system, the action $S[g_{\mu\nu}, \psi]$ in \eqref{SD} must be invariant under general diffeomorphisms. By applying an infinitesimal coordinate transformation $x^\mu \rightarrow x^\mu + \xi^\mu$, the total variation of the action must vanish
\begin{equation}
    \delta_\xi S = \int d^4x \sqrt{-g} \left[ \frac{\delta S}{\delta g^{\mu\nu}} \delta_\xi g^{\mu\nu} + \frac{\delta S}{\delta \psi} \delta_\xi \psi \right] = 0
\end{equation}
In this expression, we define $E_{\mu\nu} \propto \frac{\delta S}{\delta g^{\mu\nu}}$ as the equation of motion \eqref{eisteineq}, and $\mathcal{E}_\psi = \frac{\delta S}{\delta \psi}$ dictates the dynamics of the scalar field. We know that the variations of the fields under diffeomorphisms (Lie derivatives) are given by:
\begin{eqnarray}
    \delta_\xi g^{\mu\nu} &=& \nabla^\mu \xi^\nu + \nabla^\nu \xi^\mu,\\
    \delta_\xi \psi &=& \xi^\mu \nabla_\mu \psi.
\end{eqnarray}
Substituting these variations into the integral,  integrating by parts to remove the covariant derivatives from the arbitrary parameter $\xi^\nu$, and using $\nabla_\mu G^{\mu\nu} = 0$ we obtain the Generalized Bianchi Identity of the theory:
\begin{equation}
   \label{bianchi} \nabla^\mu \mathcal{H}_{\mu\nu} = \mathcal{E}_\psi \nabla_\nu \psi.
\end{equation}
    This is an intrinsic geometric identity. The immediate physical consequence is that when the scalar field is \textit{on-shell}, that is, when its equation of motion is satisfied ($\mathcal{E}_\psi = 0$), the divergence of the tensor perfectly vanishes
    \begin{equation}
        \nabla^\mu \mathcal{H}_{\mu\nu} = 0.
    \end{equation}
This result guarantees energy-momentum conservation in the standard EGB theory.

In coordinates, equations \eqref{eisteineq} and (\ref{r4EGBgfeq1}) are given by
\begin{align}
& \Lambda + \frac{f'}{2r} + \frac{\alpha}{2r}\Big[2 f \psi'^2 \left(3 - 2 r \psi'\right) f' 
- 2 f^2 \psi' \left(r \psi'^3 - 4 \psi'' + 4 r \psi' \psi''\right)
\Big] = T_{t}^{\;t},
\\
& \Lambda + \frac{f'}{2r} + \frac{\alpha}{2r}\Big[2 f^2 \psi'^3 \left(-4 + 3 r \psi'\right) 
+ 2 f \psi'^2 \left(3 - 2 r \psi'\right) f'
\Big] = T_{r}^{\;r},
\\
& \Lambda + \frac{f''}{2} + \alpha \Big[ 
\psi'^2 f'^2 
- f^2 \psi'^2 \big(\psi'^2 + 4 \psi''\big) 
+ f \psi' \big(-2 \psi'^2 f' 
+ 2 f' \psi'' + \psi' f''\big)
\Big] = T_{\varphi}^{\;\varphi},
\\
& \psi' ( -1 + r \psi' ) f'^2 
- 2 f^2 \psi' \left[  \psi'^2 + ( -2 + 3 r \psi' ) \psi'' \right]
+ f \left[ - f' \left( \psi'' + \psi' \left( \psi' ( -5 + 4 r \psi' ) - 2 r \psi'' \right) \right)  \psi' ( -1 + r \psi' ) f'' \right] = 0.
\end{align}
Before proceeding, we simplify the equations above. We have:
\begin{equation}
\left[
\frac{\Lambda r^2}{2} + \frac{f}{2} + \frac{\alpha}{2}\,f^2 \psi'^2 (3-2r\psi')
\right]'
=
r\,T_{t}^{\;t}
-\alpha f^2 \psi'(1-r\psi')(\psi'^2+\psi'').
\end{equation}
\begin{equation}
\left[
\frac{\Lambda r^2}{2} + \frac{f}{2}
+ \frac{\alpha}{2} f^2 \psi'^2 (3-2r\psi')
\right]'
=
r\,T_{r}^{\;r}
+ 3\alpha f^2 \psi'(1-r\psi')(\psi'^2+\psi'').
\end{equation}
\begin{equation}
\left[
\frac{\Lambda r^2}{2} + \frac{r f'}{2}
+ \alpha r f^2 \psi'^2 (1 - r\psi')
\right]'
=
r\,T_{\varphi}^{\;\varphi}
+ \alpha r f^2 \psi' (1 - r\psi')(\psi'^2+\psi'').
\end{equation}
\begin{equation}
\left[
f^2 \psi'(r\psi'-1)
\left(\ln f - 2\psi\right)'
\right]'=0.
\end{equation}
In order to construct a regular black hole solution, we begin by introducing a generic matter source whose stress-energy tensor $T_{\mu\nu}$ satisfies the spatial-temporal symmetry condition:
\begin{equation}
T_t^{\;t} = T_r^{\;r}.
\end{equation}
A highly motivated physical candidate that naturally obeys this relation is a Non-Linear Electrodynamics (NED) field. In particular, a purely eletric NED ansatz provides a robust mechanism to resolve the central singularity while strictly preserving this symmetry. 

By virtue of this property, when we subtract the $r-r$ component of the field equations from the $t-t$ component, the contributions from the generic source cancel out completely. Consequently, we obtain the exact same differential relation as in \cite{Hennigar:2020fkv}:
\begin{equation}
\frac{4\alpha f^2 \left( r\psi'^2 - \psi' \right) \left( \psi'^2 + \psi'' \right)}{r} = 0.
\end{equation}
This is a remarkable feature, as it implies that the scalar field solution 
\begin{equation}
\label{psireg}\psi = \ln(r/l)
\end{equation}
remains strictly valid. The presence of the  source does not spoil the scalar field configuration. Finally, by summing the $r-r$ and $t-t$ equations and applying the scalar field solution, the master equation governing the metric function $f(r)$ becomes:
\begin{equation}
\frac{\alpha f(r)f'(r)}{r^3} - \frac{\alpha f(r)^2}{r^4} + \frac{f'(r)}{2r} - \frac{1}{\ell^2} = T_t^{\;t}.
\label{eq_master_ned}
\end{equation}
Rewriting the above equation as a total derivative, and identifying the energy density of the fluid as $\rho = -T_t^{\;t}$, we obtain:\begin{equation}\frac{1}{r} \frac{d}{dr} \left[ \frac{\alpha f(r)^2}{2r^2} + \frac{f(r)}{2} - \frac{r^2}{2\ell^2} \right] = -\rho(r).\end{equation}Multiplying by $2r$, this expression can be cast into a more convenient differential form, which will be essential for finding the mass function of the spacetime:\begin{equation}\frac{d}{dr} \left[ \frac{\alpha f(r)^2}{r^2} + f(r) - \frac{r^2}{\ell^2} \right] = -2r \rho(r).\end{equation}

Integrating both sides of the equation with respect to $r$, we obtain an algebraic equation for the metric function:
\begin{equation}\label{eqf}\frac{\alpha}{r^2} f(r)^2 + f(r) - \frac{r^2}{\ell^2} = -C_1 -g(r),
\end{equation}
where we have defined the function $g(r)$ as:
\begin{equation}g(r) = 2 \int r \rho(r) dr.
\end{equation}
Here, $C_1$ is an integration constant. Notice that in the vacuum limit, where the matter source vanishes ($\rho = 0$), the  function $g$ conveniently reduces to zero. Rearranging Eq. \eqref{eqf} as a standard quadratic equation for $f(r)$, we have:
\begin{equation}\frac{\alpha}{r^2} f(r)^2 + f(r) - \left( \frac{r^2}{\ell^2} - g(r) -C_1 \right) = 0.
\end{equation}
Solving this quadratic equation for $f(r)$ yields two possible branches for the spacetime geometry:
\begin{equation}\label{eq:f_pm}f_{\pm}(r) = -\frac{r^2}{2\alpha} \left[ 1 \pm \sqrt{1 + \frac{4\alpha}{r^2} \left( \frac{r^2}{\ell^2} - C_1 - g(r) \right)} \right].
\end{equation}

The physical solution is the one associated with the negative sign in Eq. \eqref{eq:f_pm}. Furthermore, by analyzing the asymptotic behavior of $f(r)$ for large values of $r$, it is noteworthy that the physical meaning of the constant $C_1$ is the same as that presented in Ref. \cite{Alencar:2026yyz}; that is, 
\begin{equation}
    C_1 = K\, M_{eff} \to M_{eff} = \frac{C_1}{K}, \;\;\;\; K = \sqrt{1 + \frac{4\alpha}{\ell^2}},
\end{equation}
which implies that the Gauss–Bonnet parameter $\alpha$ must satisfy the constraint $ 0< \alpha < -\ell^2/4$.
The event horizon, defined by $f(r=r_h)=0$, is given implicitly by the equation:
\begin{equation}
    \frac{r_h^2}{\ell^2} - M_{eff}K - g(r_h) = 0.
\end{equation}

\subsection{Conditions for regular curvature and NED configuration}
The curvature properties of the spacetime can be characterized through the Kretschmann scalar,
\begin{equation}
K \equiv R_{\mu\nu\rho\sigma}R^{\mu\nu\rho\sigma}.
\end{equation}
For the class of static, circularly symmetric configurations considered here, one finds
\begin{equation}
K(r) = \big(f''(r)\big)^2 + 2\left(\frac{f'(r)}{r}\right)^2.
\end{equation}
From this expression, it is clear that potential curvature singularities may arise from divergences in either $f''(r)$ or the combination $f'(r)/r$.

We now investigate the behavior of $K(r)$ near the origin. The relevant quantity controlling regularity is
\begin{equation}
A_1(r) = 1 + \frac{4\alpha}{\ell^2} - \frac{4\alpha}{r^2}\big(C_1 + g(r)\big),
\end{equation}
where the quantity inside the square root plays a crucial role in determining the regularity of the solution.

As discussed in Ref.~\cite{Hennigar:2020fkv}, in the particular case $g(r)=0$, the spacetime is always singular at the origin. Indeed, in this limit one has
\begin{equation}
A(r) = 1 + \frac{4\alpha}{\ell^2} - \frac{4\alpha C_1}{r^2},
\end{equation}
which diverges as $r \to 0$. As a consequence, the metric function behaves as $f(r) \sim r$, leading to
\begin{equation}
K(r) \sim \frac{1}{r^2},
\end{equation}
thus signaling the presence of a curvature singularity at $r=0$.

More general, a Taylor expansion of $g(r)$ around $r=0$ leads,
\begin{equation}
g(r) = g_0 + g_1 r + g_2 r^2 + \mathcal{O}(r^3),
\end{equation}
shows that $A_1(r)$ generically develops divergent terms proportional to $1/r^2$ and $1/r$. The absence of such divergences requires the conditions
\begin{equation}
g(0) +C_1 =0, \qquad g'(0) = 0.
\end{equation}

When these constraints are satisfied, $A_1(r)$ approaches a finite constant as $r \to 0$, and the metric function behaves as $f(r) \sim r^2$. Consequently, both $f''(r)$ and $f'(r)/r$ remain finite, ensuring that the Kretschmann scalar is regular at the origin,
\begin{equation}
K(r) \to \text{const}.
\end{equation}

Conversely, if $g(0) +C_1\neq 0$, the function $A_1(r)$ diverges as $1/r^2$, leading to $f(r) \sim r$ and
\begin{equation}
K(r) \sim \frac{1}{r^2},
\end{equation}
which signals the presence of a curvature singularity at $r=0$.

\subsection{An Explicit Regular Solution and its NED Source}
As previously mentioned, in order to construct regular solutions we shall consider a nonlinear electrodynamics (NED) source. In this framework, the matter sector is described by a Lagrangian density $\mathcal{L}(F)$, where $F = F_{\mu\nu}F^{\mu\nu}$ is the electromagnetic invariant. The corresponding energy-momentum tensor is:
\begin{equation}T_{\mu\nu} = 4 \mathcal{L}_F F_{\mu\lambda} F_{\nu}^{\lambda} - g_{\mu\nu} \mathcal{L}(F),\end{equation}
where $\mathcal{L}_F \equiv d\mathcal{L}/dF$.
For a purely electric configuration, where the non-vanishing component is $F_{tr} = -F_{rt} = E(r)$, the invariant $F$ is given by $F = -2E(r)^2$.

The generalized Maxwell equation for this case is given by
\begin{equation}
   \label{max} \frac{1}{\sqrt{-g}}\partial_\mu\left(\sqrt{-g}\mathcal{L}_F F^{\mu\nu}\right) = 0 \rightarrow E(r) = \frac{q}{r\mathcal{L}_F}, 
\end{equation}
where $q$ is a constant.

The energy density $\rho = -T^t_{\;t}$ and $T_\varphi^\varphi$ is then given by:
\begin{eqnarray}
  \label{ned1}  T^t_t &=& T_r^r= -\mathcal{L} - 4E^2 \mathcal{L}_F = -\mathcal{L} - \frac{4q^2}{r^2\mathcal{L}_F},\\
 \label{ned2}    T_\varphi^\varphi &=& -\mathcal{L}
\end{eqnarray}

To ensure the regularity of the gravitational solution at the origin, we prescribe the NED contribution to the metric function as:\begin{equation}g(r) = \frac{q^2}{r^2 + a^2},\end{equation}where $q$ is the electric charge and $a$ is a regulator parameter. From the field equations, the energy density associated with this profile is obtained via $\rho(r) = -\frac{1}{2r}g'(r)$, resulting in:\begin{equation}\rho(r) = \frac{q^2}{(r^2 + a^2)^2}.\end{equation}

The field equations using \eqref{psireg} are reduced to
\begin{align}
\label{ug1coor}& \Lambda + \frac{f'}{2r} + \frac{\alpha f'f}{r^3} -\frac{\alpha f^2}{r^4} = T_{t}^{\;t},
\\
\label{ug2coor}& \Lambda + \frac{f'}{2r} + \frac{\alpha f'f}{r^3} -\frac{\alpha f^2}{r^4}= T_{r}^{\;r},
\\
& \Lambda + \frac{f''}{2} + \alpha\left[
\frac{f'^2}{r^2} + \frac{3 f^2}{r^4} - \frac{4 f f'}{r^3} + \frac{f f''}{r^2}
\right] = T_{\varphi}^{\;\varphi}.
\end{align}
The form of $\mathcal{L}(F(r))$ is obtained directly from the equation for $T_{\varphi}^{\;\varphi}$. Thus, we have
\begin{equation}
    \mathcal{L}(F(r))= -\frac{q^2 \left(a^2-3 r^2\right)}{\left(a^2+r^2\right)^3}.
\end{equation}
To find the function $\mathcal{L}_F$, it is sufficient to subtract the angular part from the temporal part ($T_\varphi^\varphi - T_t^t$), from which we obtain
\begin{equation}
    \mathcal{L}_F(r) = -\frac{(r^2 + a^2)^3}{r^4},
\end{equation}
which, using \eqref{max}, directly gives us the electric field
\begin{equation}
    E(r) = - \frac{q r^3}{(r^2 +a^2)^3}
\end{equation}
and the invariant $F(r)$
\begin{equation}
    F(r) = -\frac{2q^2r^6}{(r^2 + a^2)^6}.
\end{equation}
It is straightforward to verify that this set of functions $\{E(r), \mathcal{L}(r), \mathcal{L}_F(r)\}$ satisfies the integrability condition identically, ensuring that the model stems from a well-defined $\mathcal{L}(F)$ theory. 

In this context $2+1$, we can invert the equation for $F(r)$ to write $r(F)$ and thus express $\mathcal{L}$ as a function of the invariant $F$
\begin{equation}
    \mathcal{L}(F) = \frac{q^2}{a^4}\left(\frac{1 - \sqrt{1 - 4\mathcal{K}(F)}}{2}\right)^2(2\sqrt{1 - 4\mathcal{K}(F)}-1) ,
\end{equation}
where
\begin{equation}
    \mathcal{K}(F) = \left(\frac{a^6|F|}{2q^2}\right)^{1/3}.
\end{equation}
At the origin, the metric function $g(r)$ and its derivatives behave as:\begin{equation}g(0) = \frac{q^2}{a^2}, \quad g'(0) = 0, \quad g''(0) = -\frac{2q^2}{a^4}.\end{equation}This confirms that the NED contribution is smooth and finite, providing a regular core. Substituting $g(r)$ into the general solution, the metric function $f(r)$ becomes:\begin{equation}f(r) = -\frac{r^2}{2\alpha} \left[ 1 - \sqrt{ 1 + \frac{4\alpha}{r^2} \left( \frac{r^2}{\ell^2} - C_1 - \frac{q^2}{r^2 + a^2} \right) } \right].\end{equation}The horizon condition $f(r_h) = 0$ then reduces to:\begin{equation}\frac{r_h^2}{\ell^2} - M_{\text{eff}}K - \frac{q^2}{r_h^2 + a^2} = 0.\end{equation}

By rearranging this expression into a polynomial form, we find a biquadratic equation for the horizon radius, namely $r_h^4 + (a^2 - M_{\text{eff}}K\ell^2)r_h^2 - (M_{\text{eff}}K\ell^2 a^2 + q^2\ell^2) = 0$. Crucially, since the constant term is strictly negative (assuming $M, q, a, \ell > 0$), Descartes' Rule of Signs guarantees the existence of exactly one positive real root for $r_h^2$. Consequently, the system possesses a single event horizon located at:\begin{equation}r_h = \sqrt{\frac{M_{\text{eff}}K\ell^2 - a^2 + \sqrt{(a^2 - M_{\text{eff}}K\ell^2)^2 + 4(M_{\text{eff}}K\ell^2 a^2 + q^2\ell^2)}}{2}}.\end{equation}From a physical perspective, the absence of a second root implies that this geometry does not admit a Cauchy horizon, a feature that distinguishes it from the standard Reissner-Nordström-like solutions. Furthermore, this single-horizon structure ensures a simplified causal global topology while maintaining the asymptotic Anti-de Sitter behavior. Such a configuration avoids the inner-horizon instabilities typically associated with charged black holes, providing a more robust background for thermodynamic and holographic analyses.

\section{Black Bounce Geometries in Lower-Dimensional Gauss–Bonnet Gravity}
In this section, we will consider the action (\ref{SD}) and spherical symmetry, with metric 
\begin{equation}
    \label{bbmetric}ds^2=-f(r)dt^2+\frac{1}{f(r)}dr^2+\Sigma(r)^2d\varphi^2.
\end{equation}

For $\Sigma = r$, this metric reduces to the well-known 3-dimensional black hole metric \eqref{bh3}. For a general $\Sigma$ function, this metric represents a black bounce in 2+1 dimensions.

In coordinates, equations \eqref{eisteineq} and (\ref{r4EGBgfeq1}) are given by
\begin{align}
& \Lambda + \frac{f' \Sigma'}{2 \Sigma} + \frac{f \Sigma''}{\Sigma} + \alpha \left( -2 f \psi'^3 f' + \frac{3 f \psi'^2 f' \Sigma'}{\Sigma} - f^2 \psi'^2 (\psi'^2 + 4 \psi'') + \frac{2 f^2 \psi' (2 \Sigma' \psi'' + \psi' \Sigma'')}{\Sigma} \right) = T_{t}^{\;t} \label{rhobb}, \\
& \Lambda + \frac{f'\Sigma'}{2\Sigma}+ \frac{\alpha \left( 6 f^2 \Sigma \psi'^4 - 4 f \Sigma \psi'^3 f' - 8 f^2 \psi'^3 \Sigma' + 6 f \psi'^2 f' \Sigma' \right)}{2 \Sigma} = T_{r}^{\;r} , \\
&\Lambda + \frac{f''}{2} + \alpha \left( \psi'^2 f'^2 - f^2 \psi'^2 \left( \psi'^2 + 4\psi'' \right) + f \psi' \left( -2\psi'^2 f' + 2f'\psi'' + \psi' f'' \right) \right) = T_{\varphi}^{\;\varphi}
, \label{ptbb}\\
& \begin{aligned}\label{scalarbb}
-\frac{1}{2 \Sigma} \Big( \psi' f'^2 (\Sigma \psi' + \Sigma') &+ f \left( \Sigma \psi' (2 f' (2 \psi'^2 + \psi'') + \psi' f'') + \Sigma' (f' (5 \psi'^2 + \psi'') + \psi' f'') + \psi' f' \Sigma'' \right) \\
&+ 2 f^2 \psi' \left( \psi'^2 \Sigma' + 2 \Sigma' \psi'' + \psi' (3 \Sigma \psi'' + \Sigma'') \right) \Big) = 0.
\end{aligned}
\end{align}
which can be simplified as
\begin{align}
\label{ttbbs}\left[
\frac{\Lambda}{2}\Sigma^2
+\frac{f}{2}(\Sigma')^2
+\frac{\alpha}{2}f^2\psi'^2
\left(3(\Sigma')^2-2\Sigma\Sigma'\psi'\right)
\right]'
&=
\Sigma\Sigma' T_t^{\;t}
+\alpha f^2\psi'(\Sigma\psi'-\Sigma')
\Sigma'(\psi'^2+\psi'') \\
&\quad
-\Sigma''\left[
\alpha f^2\psi'^2(\Sigma\psi'-\Sigma')
\right].
\end{align}
\begin{align}
\left[
\frac{\Lambda}{2}\Sigma^2
+\frac{f}{2}(\Sigma')^2
+\frac{\alpha}{2}f^2\psi'^2
\left(3(\Sigma')^2-2\Sigma\Sigma'\psi'\right)
\right]'
&=
\Sigma\Sigma' T_r^{\;r}
+3\alpha f^2\psi'\Sigma'(\Sigma'-\Sigma\psi')(\psi'^2+\psi'') \\
&\quad
+\Sigma''\left[
f\Sigma'
+\alpha f^2\psi'^2(3\Sigma'-\Sigma\psi')
\right].
\end{align}
\begin{align}
\label{ppbbs}\left[
\frac{\Lambda}{2}\Sigma^2
+\frac{1}{2}\Sigma\Sigma' f'
+\frac{\alpha}{2}\Sigma^2 f^2\psi'^2(1-\Sigma\psi')
\right]'
&=
\Sigma\Sigma' T_\varphi^\varphi
+\alpha \Sigma^2 f^2 \psi'(1-\Sigma\psi')(\psi'^2+\psi'') \\
&\quad
+\Sigma''\left[
\frac{1}{2}\Sigma f'
+\alpha \Sigma^2 f^2\psi'^2(1-\Sigma\psi')
\right].
\end{align}
\begin{equation}
\label{psibb}\left[
f^2\psi'(\Sigma\psi'-\Sigma')
\left(\ln f-2\psi\right)'
\right]'=0.
\end{equation}
the general solution of equation \eqref{psibb} for the black bounce case is given by 
\begin{equation}
    f^2\psi'\left(\psi'-\frac{\Sigma'}{\Sigma}\right)
=
\frac{C}{\Sigma\left(\ln f-2\psi\right)'}.
\end{equation}
If we consider $C=0$, the last equation can be solved for $\psi$. In order to recover the case $\Sigma=r$ we must have 
\begin{equation}
  \label{solpsibb}  \psi=\ln (\Sigma/l).
\end{equation}

This black bounce case is mathematically much more complicated than the black hole case, as the fact that $\Sigma' \neq 1$ implies a non-trivial coupling between the scalar field sector and the gravitational sector. 

Now, to proceed, as previously stated, one strategy is to fix the geometry through the function $f(r)$ and determine the fields that generate it. In this case, we will adopt a procedure analogous to Simpson-Visser and consider a modified version for the function $f$ where, in the vacuum solution function for $f(r)$ given in \cite{Hennigar:2020fkv}, we will make the substitution $r \to \Sigma(r)$. Thus, we will consider
\begin{equation}
\label{fbb}f(r) = -\frac{\Sigma^2}{2\alpha} \left[ 1 - \sqrt{1 +\frac{4\alpha}{\Sigma^2} \left( \frac{\Sigma^2}{\ell^2} - C_1 \right)} \right].
\end{equation}

\subsection{Conditions for regular curvature }\label{nedbb}

As discussed in the previous section, the curvature properties of the spacetime are characterized by the Kretschmann scalar, $K \equiv R_{\mu\nu\rho\sigma}R^{\mu\nu\rho\sigma}$. For the class of static, circularly symmetric metrics given by \eqref{bbmetric}, the Kretschmann scalar is explicitly expressed as:
\begin{equation}
    K(r) = \big(f''(r)\big)^2 + 2\left(\frac{f'(r)\Sigma'(r)}{\Sigma(r)}\right)^2 + 2\left(\frac{f(r)\Sigma''(r)}{\Sigma(r)}\right)^2.
\end{equation}
From this expression, it is evident that the regularity of the origin depends on the behavior of $f(r)$ and $\Sigma(r)$ as $r \to 0$. In standard singular configurations where $\Sigma(r) = r$, the $1/r$ terms typically lead to divergences.

The regularity of the metric function is determined by the quantity:
\begin{equation}
    A_2(r) = 1 + \frac{4\alpha}{\ell^2} - \frac{4\alpha C_1}{\Sigma(r)^2},
\end{equation}
where the metric function is defined as
\begin{equation}
    f(r) = -\frac{\Sigma(r)^2}{2\alpha} \left[ 1 - \sqrt{A_2(r)} \right].
\end{equation}
We now impose the conditions for a regular, non-vanishing core at the origin:
\begin{equation}
    \Sigma(0) = a, \quad \text{and} \quad \Sigma'(0) = 0,
\end{equation}
where $a$ is a non-zero constant. Under these conditions, the argument $A_2(r)$ remains strictly finite at the origin:
\begin{equation}
    A_2(0) = 1 + \frac{4\alpha}{\ell^2} - \frac{4\alpha C_1}{a^2}.
\end{equation}
Consequently, provided the derivatives $\Sigma''(0)$ and $f''(0)$ are finite, as required for any smooth manifold, the behavior of the individual terms in $K(r)$ is as follows:

\begin{itemize}
    \item \textbf{First term:} $\big(f''(r)\big)^2$ is finite by the assumption of smoothness.
    \item \textbf{Second term:} The ratio $\frac{f'(r)\Sigma'(r)}{\Sigma(r)}$ vanishes at the origin. Since $\Sigma'(0)=0$ and $f'(0)=0$ (as $f$ is a function of $\Sigma$), the numerator vanishes while the denominator $\Sigma(0)=a$ remains non-zero.
    \item \textbf{Third term:} The ratio $\frac{f(r)\Sigma''(r)}{\Sigma(r)}$ is finite, approaching the value $f(0)\Sigma''(0)/a$.
\end{itemize}

Since all terms in the Kretschmann scalar remain bounded, we conclude that:
\begin{equation}
    K(r) \xrightarrow{r \to 0} \text{constant}.
\end{equation}
Therefore, the requirements that the angular radius is a non-zero constant at the origin ($\Sigma(0)=a$) and its profile is smooth ($\Sigma'(0)=0$) are sufficient to ensure that the spacetime is free of curvature singularities.

\subsection{An Explicit Simpson–Visser Black Bounce and its NED–Scalar Sources}

To find the sources that support this geometry, we consider now that the total energy-momentum tensor is composed of two contributions,
\begin{equation}
T_{\mu\nu}^{\text{(tot)}} = T_{\mu\nu}^{(\phi)} + T_{\mu\nu}^{(\text{NED})},
\end{equation}
where $T_{\mu\nu}^{(\phi)}$ corresponds to the energy momentum-tensor of an additional scalar field $\phi$ and $T_{\mu\nu}^{(\text{NED})}$ to a NED sector.

For the class of NED models under consideration, one has
\begin{equation}
T^{t}_{\ t}\big|_{\text{NED}} = T^{r}_{\ r}\big|_{\text{NED}},
\end{equation}
so that their contribution cancels out in the combination
$T^{t}_{\ t} - T^{r}_{\ r}$. Therefore,
\begin{equation}
T^{t}_{\ t} - T^{r}_{\ r} = 
T^{t}_{\ t}\big|_{\phi} - T^{r}_{\ r}\big|_{\phi}.
\end{equation}

Starting from the geometric identity obtained by subtracting the $(t,t)$ and $(r,r)$ components of the field equations, we have
\begin{equation}
\frac{\Sigma''}{\Sigma}\left(f+2\alpha f^2\psi'^2\right)
-4\alpha f^2\psi'(\psi'^2+\psi'')
\left(\psi'-\frac{\Sigma'}{\Sigma}\right)
= T^{t}_{\ t} - T^{r}_{\ r}.
\end{equation}

Now, using \eqref{solpsibb}, the second term vanishes identically, and the equation reduces to
\begin{equation}
\frac{\Sigma''}{\Sigma}
\left(
f + 2\alpha f^2 \frac{(\Sigma')^2}{\Sigma^2}
\right)
= T^{t}_{\ t} - T^{r}_{\ r}.
\end{equation}

We now model the remaining matter sector as a scalar field with Lagrangian
\begin{equation}
\mathcal{L} = -\frac{h(\phi)}{2}g^{\mu\nu}\partial_\mu\phi\partial_\nu\phi - V(\varphi),
\end{equation}
where $h(\phi)>0$ corresponds to a canonical scalar field and $h(\phi)<0$ to a phantom field. The function $h(\phi)$ modulates the kinetic contribution, and is widely used in the literature as an alternative to completely ghost or canonical scalar fields, being known as partly phantom scalar fields \cite{Silva:2025fqj,Crispim:2024lzf,Bronnikov:2022bud}.  The energy-momentum tensor is
\begin{equation}
T_{\mu\nu} = h(\phi)\partial_\mu\phi\partial_\nu\phi - g_{\mu\nu}\left(\frac{1}{2}h(\phi)\partial_\alpha\phi\partial^\alpha\phi + V(\phi)\right).
\end{equation}
For a static configuration $\phi = \phi(r)$, one obtains
\begin{equation}
T^{t}_{\ t} - T^{r}_{\ r} = -h(\phi(r)) f \phi'^2.
\end{equation}

Therefore, the field equation becomes
\begin{equation}
\Sigma''
\left(
f + 2\alpha f^2 \frac{(\Sigma')^2}{\Sigma^2}
\right)
=
-h(\phi(r)) f\Sigma \phi'^2.
\end{equation}

Finally, isolating $h\phi'(r)^2$, we obtain
\begin{equation}
h(\phi(r))\phi'^2 =-
\frac{\Sigma''}{\Sigma}
\left(
1 + 2\alpha f \frac{(\Sigma')^2}{\Sigma^2}
\right) = -\frac{\Sigma''}{\Sigma} \left[ 1 - (\Sigma')^2 \left( 1 - \sqrt{1 + \frac{4\alpha}{\ell^2} - \frac{4\alpha C_1}{\Sigma^2}} \right) \right].
\end{equation}

We can now elegantly consider that the function $h(\phi)$ is given by
\begin{equation}
    h(\phi(r))= - \left[ 1 - (\Sigma')^2 \left( 1 - \sqrt{1 + \frac{4\alpha}{\ell^2} - \frac{4\alpha C_1}{\Sigma^2}} \right) \right],
\end{equation}
what does it imply
\begin{equation}
    \phi'^2 = \frac{\Sigma''}{\Sigma}.
\end{equation}

For a Simpson-Visser-like scenario, we have $\Sigma(r) = \sqrt{r^2 + a^2}$, which implies that the scalar field $\phi$ is given by the known arctangent form:
\begin{equation}
    \phi(r) = \phi_0 + \arctan\left(\frac{r}{a}\right),
\end{equation}
where $\phi_0$ is a constant of integration that we can consider to be zero.

Therefore, we can express $h$ as a function of $r$ and as a function of the field $\phi$:
\begin{eqnarray}
h(\phi(r)) &=& -1  + \frac{r^2}{r^2+a^2} \left( 1 - \sqrt{1 + \frac{4\alpha}{\ell^2} - \frac{4\alpha C_1}{r^2+a^2}} \right) ,\\
h(\phi) &=& - 1 + \sin^2\phi \left( 1 - \sqrt{1 + \frac{4\alpha}{\ell^2} - \frac{4\alpha C_1 \cos^2\phi}{a^2}} \right) .
\end{eqnarray}

The sign analysis shows that at the throat ($r=0$), $h(0) = -1$. To evaluate its behavior elsewhere, we can write the function as $h(\phi) = -1 + \sin^2\phi \, \mathcal{A}(\phi)$, where $\mathcal{A}(\phi)$ represents the term in parentheses. As $r \to \infty$ (implying $\sin^2\phi \to 1$), the function asymptotically approaches $-\sqrt{1+4\alpha/\ell^2}$. For a well-defined metric where the geometry remains real-valued, $\mathcal{A}(\phi)$ is bounded such that the product $\sin^2\phi \, \mathcal{A}(\phi)$ can never overcome the initial $-1$ term. In fact, at spatial infinity, the positivity of $4\alpha/\ell^2$ ensures that $\mathcal{A}(\phi)$ becomes strictly negative. Consequently, $h$ remains strictly negative ($h < 0$) throughout the entire spacetime, meaning that the scalar field is consistently a ghost field.

Using the equation of motion for the scalar field (Klein-Gordon equation):
\begin{equation}
\frac{1}{\sqrt{-g}} \partial_\mu \left( \sqrt{-g} h(\phi) g^{\mu\nu} \partial_\nu \phi \right) - \frac{1}{2} \frac{dh}{d\phi} g^{\mu\nu} \partial_\mu \phi \partial_\nu \phi - \frac{dV}{d\phi} = 0,
\end{equation}
we have
\begin{equation}
    \frac{1}{\Sigma} \frac{d}{dr} \left( \Sigma f(r) h(\phi) \phi' \right) - \frac{1}{2} \frac{dh}{d\phi} f(r) \phi'^2 - \frac{dV}{d\phi} = 0.
\end{equation}

Given the relation $\phi'^2 = \Sigma''/\Sigma$, we can express the derivative of the potential as:
\begin{equation}
   \label{Vr} \frac{dV}{dr} = \frac{\phi'(r)}{\Sigma(r)} \frac{d}{dr} \left( \Sigma(r) f(r) h(r) \phi'(r) \right) - \frac{1}{2} f(r) \phi'(r)^2 \frac{dh}{dr}.
\end{equation}

Integrating with respect to $r$ (or $\phi$), the potential $V(\phi)$ is determined up to a quadrature:
\begin{equation}
   \label{vp} V(\phi) = \int \left[ \frac{1}{\Sigma} \frac{d}{dr} \left( \Sigma f h \phi' \right) - \frac{1}{2} f \phi'^2 \frac{dh}{dr} \right] \frac{dr}{d\phi} d\phi.
\end{equation}

For the Simpson-Visser metric, where $\Sigma = a \sec \phi$ and $\phi' = \frac{\cos^2 \phi}{a}$, all terms inside the integral are explicit functions of $\phi$ through $h(\phi)$ and $f(\phi)$. The expression for the potential, although analytical, is extremely large. Because of this, it will not be written here, but can be found in the appendix for further reference.

Before proceeding with our analysis, we note that the equations \eqref{ttbbs}-\eqref{ppbbs} can be written, using the solution for the scalar field $\psi$ as
\begin{eqnarray}
   && \left(\frac{\Lambda\Sigma^2}{2} + \frac{f\Sigma'^2}{2} + \frac{\alpha f^2\Sigma'^4}{2\Sigma^2}\right)' = \Sigma'\Sigma T_t^t, \\
  &&  \left(\frac{\Lambda\Sigma^2}{2} + \frac{f\Sigma'^2}{2} + \frac{\alpha f^2\Sigma'^4}{2\Sigma^2}\right)' = \Sigma'\Sigma T_r^r + f\Sigma''\Sigma'\left(1 + \frac{2\alpha f\Sigma'^2}{\Sigma^2}\right),\\
  && \left(\frac{\Lambda\Sigma^2}{2} + \frac{\Sigma\Sigma'f'}{2} + \frac{\alpha f^2\Sigma'^2}{2}(1 - \Sigma')\right)' = \Sigma'\Sigma T_\varphi^\varphi + \alpha f^2\Sigma'\Sigma''(1 - \Sigma'^2) + \frac{\Sigma\Sigma''f'}{2},
\end{eqnarray}
so that the difference $   T_t^t - T_\varphi^\varphi$ is given by
\begin{equation}
    T_t^t - T_\varphi^\varphi = \frac{1}{\Sigma\Sigma'}\left(\frac{f\Sigma'^2}{2} - \frac{\Sigma\Sigma' f'}{2} + \frac{\alpha f^2}{2}\left[\frac{\Sigma'^4}{\Sigma^2} - \Sigma'^2(1 - \Sigma')\right]\right)' + \frac{\Sigma''f'}{2\Sigma'}+\alpha f^2\frac{\Sigma''}{\Sigma}(1 - \Sigma'^2)
\end{equation}

As previously mentioned, the additional scalar field $\phi$ possesses symmetry ${T^{t}_{\ t}}^{(\phi)} = {T^{\varphi}_{\ \varphi}}^{(\phi)}$. Consequently, the right-hand side is entirely determined by the nonlinear electrodynamics sector,
\begin{equation}
T^{t}_{\ t} - T^{\varphi}_{\ \varphi}
=
T^{t}_{\ t}\big|_{\text{NED}} - T^{\varphi}_{\ \varphi}\big|_{\text{NED}}.
\end{equation}

For a general nonlinear electrodynamics described by a Lagrangian $\mathcal{L}(F)$, where $F = F_{\mu\nu}F^{\mu\nu}$, the energy-momentum tensor is given by \eqref{ned1}-\eqref{ned2}. The generalized Maxwell equation for this black bounce case is given by
\begin{equation}
   \label{max} \frac{1}{\sqrt{-g}}\partial_\mu\left(\sqrt{-g}\mathcal{L}_F F^{\mu\nu}\right) = 0 \rightarrow E(r) = \frac{q}{\Sigma\mathcal{L}_F}, 
\end{equation}
where $q$ is a constant.

Finally, the master equation governing the nonlinear electrodynamics sector reads
\begin{equation}
    \frac{4q^2}{\Sigma^2\mathcal{L}_F} = \frac{1}{\Sigma\Sigma'}\left(\frac{f\Sigma'^2}{2} - \frac{\Sigma\Sigma' f'}{2} + \frac{\alpha f^2}{2}\left[\frac{\Sigma'^4}{\Sigma^2} - \Sigma'^2(1 - \Sigma')\right]\right)' + \frac{\Sigma''f'}{2\Sigma'}+\alpha f^2\frac{\Sigma''}{\Sigma}(1 - \Sigma'^2)
\end{equation}
The Lagrangian follows directly from the angular component
\begin{equation}
   \mathcal{L}(r) = \frac{1}{\Sigma'\Sigma}\left(\frac{\Lambda\Sigma^2}{2} + \frac{\Sigma\Sigma'f'}{2} + \frac{\alpha f^2\Sigma'^2}{2}(1 - \Sigma')\right)' - \frac{\Sigma''f'}{2\Sigma'}-\alpha f^2\frac{\Sigma''}{\Sigma}(1 - \Sigma'^2)
\end{equation}

Although we can write these equations explicitly in terms of $r$, their expressions are rather lengthy and will be presented in the appendix for reference. For the electric field in this case, we can write it explicitly as:
\begin{equation}
\begin{split}
    E(r) &= \frac{\sqrt{a^2+r^2}}{32 q} \Biggl\{ 
    -\frac{8 C_1 r^2 \left[r^2+r (a^2+r^2)^{3/2}-(a^2+r^2)^2\right] (1-\sqrt{\Delta})}{(a^2+r^2)^3 \sqrt{\Delta}} \\
    &\quad + \frac{4 (\sqrt{\Delta}-1)}{\alpha} - \frac{4 r^2 (\sqrt{\Delta}-1)}{(a^2+r^2) \alpha} + \frac{2 a^4 (\sqrt{\Delta}-1)^2}{(a^2+r^2) \alpha} \\
    &\quad + \frac{4 r^2 \left[r^2+r (a^2+r^2)^{3/2}-(a^2+r^2)^2\right] (\sqrt{\Delta}-1)^2}{(a^2+r^2)^2 \alpha} \\
    &\quad + \frac{(\sqrt{\Delta}-1)^2}{(a^2+r^2)^{5/2} \alpha} \Bigl[ -2 r^4 \sqrt{a^2+r^2} + a^6 \left(3 r-2 \sqrt{a^2+r^2}\right) \\
    &\qquad\qquad + 2 a^4 r^2 \left(3 r-2 \sqrt{a^2+r^2}\right) \\
    &\qquad\qquad + a^2 \left(3 r^5+4 r^2 \sqrt{a^2+r^2}-2 r^4 \sqrt{a^2+r^2}\right) \Bigr] \\
    &\quad + 4 \left[ \frac{8 C_1^2 r^2 \alpha}{(a^2+r^2)^3 \Delta^{3/2}} - \frac{2 C_1}{(a^2+r^2) \sqrt{\Delta}} + \frac{1-\sqrt{\Delta}}{\alpha} \right] \Biggr\},
\end{split}
\end{equation}
where
\begin{equation}
\Delta \equiv
1+\frac{4\alpha}{\ell^2}
-\frac{4 C_1 \alpha}{a^2+r^2}.
\end{equation}

We can analyze the asymptotic behavior of the electric field near the origin, from where we obtain
\begin{eqnarray}
     E(r) &\to&  -\frac{C_1}{4aq\sqrt{\Delta(0)}}.
\end{eqnarray}

We note that the electric field is regular and constant at the origin.

\section{Regular Maxwell-Supported Solutions in Unimodular Gauss–Bonnet Gravity}

In the previous sections, we followed the phenomenological prescription widely adopted in the current literature of employing NED to construct geometrically regular black hole solutions. However, as discussed in the literature (Refs.\cite{Alencar:2026qeb,Novello:1999pg,Russo:2022qvz,deMelo:2014isa,Russo:2024llm,Abalos:2015gha,dePaula:2024yzy}), although NED is a powerful tool for obtaining regular configurations, it may introduce potential pathologies, such as issues related to birefringence and the loss of a well-defined causal structure for electromagnetic wave propagation (often associated with effective metrics and superluminal modes).

On the other hand, the framework of unimodular gravity has recently been explored as an elegant alternative for deriving regular solutions sourced by standard Maxwell gauge fields (Refs.\cite{Alencar:2026hjf,Alencar:2026oxy,3137491}), in four-dimensional spacetimes.

In this sense, in the present subsection we adopt this framework in order to circumvent the reliance on NED employed in the previous section.
We begin by briefly recalling how unimodular gravity modifies the standard Einstein equations. In this framework, the metric determinant is fixed, $\sqrt{-g}=\text{const}$, which can be implemented through a Lagrange multiplier $\Lambda(x)$. Starting from the action
\begin{equation}
S_{UG} = \int d^D x \left[ \sqrt{-g}\, R + \Lambda(x)\,(\sqrt{-g}-1) + \mathcal{L}_M \right],
\end{equation}
variation with respect to the metric yields
\begin{equation}
G_{\mu\nu} + \Lambda(x)\, g_{\mu\nu} = T_{\mu\nu}.
\end{equation}
Taking the covariant divergence and using $\nabla_\mu G^{\mu\nu}=0$, one obtains
\begin{equation}
\nabla_\mu T^{\mu\nu} =  \nabla^\nu \Lambda(x).
\end{equation}
Therefore, the usual conservation law is recovered only if $\Lambda(x)$ is constant, in which case it plays the role of an effective cosmological constant arising as an integration constant.

We now recall the scalar--Gauss--Bonnet framework considered previously. The corresponding action is given by
\begin{equation}
S_{EGB} = \int d^D x \sqrt{-g} \Bigl[
R - 2\Lambda 
+ \alpha \mathcal{L}_{GB}(\psi,g)
+ \mathcal{L}_M
\Bigr],
\end{equation}
where $\mathcal{L}_{GB}(\psi,g)$ denotes the scalar Gauss--Bonnet sector. Variation with respect to the metric leads to equation \eqref{eisteineq}.
In this setup, the tensor $\mathcal{H}_{\mu\nu}$ is not identically divergence-free, without taking into account the scalar field equation for $\psi$ given in \eqref{r4EGBgfeq1}. Instead, a direct computation shows that it satisfies the identity \eqref{bianchi}. As a consequence, the generalized Bianchi identity is satisfied only on-shell.

Motivated by these considerations, we now construct the unimodular extension of the scalar--Gauss--Bonnet theory (UGB) by promoting the cosmological term to a Lagrange multiplier. The action is given by
\begin{equation}
\label{UGB_action}
S_{UGB} = \int d^D x \left\{
\sqrt{-g}\left[
R 
+ \alpha \mathcal{L}_{GB}(\psi,g)
+ \mathcal{L}_M
\right]
+ \Lambda(x)\,(\sqrt{-g}-1)
\right\}.
\end{equation}

In this framework with Lagrange multipliers, diffeomorphism symmetry can be preserved \cite{Henneaux:1989zc,Kuchar:1991xd,Alvarez:2014qca,Bufalo:2015wda}. Varying this action with respect to the metric, taking the covariant divergence and using both $\nabla_\mu G^{\mu\nu}=0$ and the on-shell condition $\nabla_\mu \mathcal{H}^{\mu\nu}=0$, we find
\begin{eqnarray}
   \label{ug1}&& G_{\mu\nu} - \alpha \mathcal{H}_{\mu\nu} + \Lambda(x)\, g_{\mu\nu}
=  T_{\mu\nu},\\
\label{ug2}&&\nabla_\mu T^{\mu\nu} =  \nabla^\nu \Lambda(x).
\end{eqnarray}

Therefore, the non-conservation of the energy--momentum tensor is controlled by the spacetime dependence of $\Lambda(x)$. In the limit where $\nabla^\nu \Lambda(x)=0$, the standard conservation law is recovered and the theory reduces to the usual Einstein--Gauss--Bonnet model with an effective cosmological constant.

The equation for the scalar field $\psi$ is not modified. Therefore, its solution remains the same within this framework. Thus, Eq.~\eqref{ug1} in coordinates, assuming $\Lambda = \Lambda(r)$ and the logarithmic solution for the field $\psi$, becomes
\begin{align}
\label{ug1coor}& \Lambda(r) + \frac{f'}{2r} + \frac{\alpha f'f}{r^3} -\frac{\alpha f^2}{r^4} = T_{t}^{\;t},
\\
\label{ug2coor}& \Lambda(r) + \frac{f'}{2r} + \frac{\alpha f'f}{r^3} -\frac{\alpha f^2}{r^4}= T_{r}^{\;r},
\\
& \Lambda(r) + \frac{f''}{2} + \alpha\left[
\frac{f'^2}{r^2} + \frac{3 f^2}{r^4} - \frac{4 f f'}{r^3} + \frac{f f''}{r^2}
\right] = T_{\varphi}^{\;\varphi}.
\end{align}

We now determine the electromagnetic field, the effective cosmological term, and the induced current directly from the field equations.

For the standard Maxwell Lagrangian
\begin{equation}
\mathcal{L}(F) = -\frac{F}{4}, \qquad F \equiv F_{\mu\nu}F^{\mu\nu},
\end{equation}
and a purely electric configuration $F_{tr}=E(r)$, one finds $
F = -2E^2$.

The energy--momentum tensor then yields
\begin{equation}
T^t_{\;t} = T^r_{\;r} = -\frac{E^2}{2},
\qquad
T^\varphi_{\;\varphi} = +\frac{E^2}{2}.
\end{equation}

Defining
\begin{eqnarray}
    \Pi_1(r) &\equiv& \frac{f'}{2r} + \frac{\alpha f'f}{r^3} - \frac{\alpha f^2}{r^4},\\
    \Pi_2(r) &\equiv& \frac{f''}{2} + \alpha\left[
\frac{f'^2}{r^2} + \frac{3 f^2}{r^4} - \frac{4 f f'}{r^3} + \frac{f f''}{r^2}
\right],
\end{eqnarray}
the field equations reduce to
\begin{align}
\Lambda(r) + \Pi_1(r) &= -\frac{E^2(r)}{2}, \\
\Lambda(r) + \Pi_2(r) &= +\frac{E^2(r)}{2}.
\end{align}

Subtracting these equations, we obtain the electric field purely in terms of the metric function:
\begin{equation}
E^2(r) = \Pi_2(r) - \Pi_1(r).
\end{equation}

Explicitly,
\begin{equation}
E^2(r) =
\frac{f''}{2}
-\frac{f'}{2r}
+\alpha\left[
\frac{f'^2}{r^2}
+\frac{4f^2}{r^4}
-\frac{5ff'}{r^3}
+\frac{ff''}{r^2}
\right].
\end{equation}

Adding the equations instead, one finds the effective cosmological function:
\begin{equation}
\Lambda(r) = -\frac{\Pi_1(r)+\Pi_2(r)}{2}.
\end{equation}

Explicitly,
\begin{equation}
\Lambda(r) =
-\frac{1}{2}\left[
\frac{f'}{2r}
+\frac{f''}{2}
+\alpha\left(
\frac{f'^2}{r^2}
+\frac{2f^2}{r^4}
-\frac{3ff'}{r^3}
+\frac{ff''}{r^2}
\right)
\right].
\end{equation}

In the unimodular framework, the non-conservation law is given by \eqref{ug2}.
For the Maxwell field, this implies
\begin{equation}
F_{\nu\alpha} J^\alpha = -\partial_\nu \Lambda(r).
\end{equation}

For a purely radial electric field, the only non-vanishing component is
\begin{equation}
J^t(r) = \frac{\Lambda'(r)}{E(r)}, \qquad
J^r = J^\varphi = 0.
\end{equation}

This completes the reconstruction of the electromagnetic sector entirely in terms of the metric function $f(r)$. For $f(r)$ given by \eqref{eq:f_pm}, we have:
\begin{align}
E^2(r) &= \frac{1}{2}\left(-\frac{g'(r)}{r} + g''(r)\right), \\[0.2cm]
\Lambda(r) &= -\frac{1}{l^2} - \frac{g'(r)}{4r} - \frac{g''(r)}{4}, \\[0.2cm]
J^t(r) &=  \frac{g'(r) - r\left(g''(r) + r g^{(3)}(r)\right)}
{2\sqrt{2}\, r^2 \sqrt{-\frac{g'(r)}{r} + g''(r)}}.
\end{align}

 More generally, since the effective energy density of the fluid is related to the metric function by $\rho(r) = g'(r)/(2r)$, the electric field can be rewritten exactly as:
\begin{equation}
    E^2(r) = r\,\rho'(r).
\end{equation}

Therefore, the requirement of a real electric field, $E^2(r) > 0$, imposes the strict condition
\begin{equation}
    \rho'(r) > 0,
\end{equation}
globally in the radial domain. In standard general relativity, this condition is deemed unphysical because isolated localized sources typically require the energy density to monotonically decrease towards spatial infinity ($\rho'(r) < 0$).

However, in the present unimodular framework, the requirement $\rho'(r) > 0$ does not constitute a pathology. The non-conservation law $\nabla_\mu T^{\mu\nu} = \nabla^\nu \Lambda(r)$ dictates that the background is dynamically active. Rather than describing an isolated charge in an empty vacuum, the theory supports configurations exchanging energy with the cosmological background.

To satisfy the geometric regularity conditions while strictly guaranteeing a positive definite energy density ($\rho > 0$) and a real electromagnetic invariant, we propose the following well-behaved geometric function:
\begin{equation}
    g(r) = \rho_0 r^2 + \frac{q^2}{r^2+a^2},
\end{equation}
where $q$ acts as the effective charge parameter, $a$ as the regularity cutoff, and $\rho_0$ is a constant representing the background vacuum energy density. Note that any purely constant term required to fix specific boundary conditions for $g(r)$ can be trivially absorbed into the integration constant $C_1$, which defines the effective mass of the spacetime.

By inserting this function into the reconstructed UGB equations, we obtain the complete exact profile for the Maxwell sector, the associated effective density $\rho(r)$, and the unimodular effective terms:
\begin{align}
    E(r) &= \frac{2q r}{(r^2+a^2)^{3/2}}, \\[0.2cm]
    \rho(r) &= \rho_0 - \frac{q^2}{(r^2+a^2)^2}, \\[0.2cm]
    \Lambda(r) &= -\left(\frac{1}{\ell^2} + \rho_0\right) - \frac{q^2(r^2-a^2)}{(r^2+a^2)^3}, \\[0.2cm]
    J^t(r) &= \frac{2q(r^2-2a^2)}{(r^2+a^2)^{5/2}}.
\end{align}

This exact solution admits a compelling physical interpretation that circumvents standard no-go paradigms. Provided that the background density satisfies $\rho_0 \ge q^2/a^4$, the energy density $\rho(r)$ remains strictly positive everywhere, avoiding violations of the weak energy condition. The resulting electric field is regular everywhere, vanishing at the origin and decaying appropriately at large distances, perfectly mimicking a localized field. 

The explicit form of $\rho(r)$ reveals that the energy density strictly grows outwards ($\rho'(r) > 0$), asymptotically approaching the vacuum density $\rho_0$. This unconventional behavior is directly sustained by the unimodular background. The effective cosmological parameter $\Lambda(r)$ acts as a dynamic reservoir, where $\rho_0$ naturally redefines the effective AdS radius of the spacetime. Its spatial gradient explicitly induces the localized current $J^t(r)$, which continuously exchanges energy with the Maxwell sector. Consequently, this configuration represents a regular central defect embedded within, and actively supported by, a pervasive cosmological background, demonstrating that unimodular gravity successfully supports regular black hole solutions sourced by standard Maxwell fields.

\section{Thermodynamics}
To initiate the thermodynamic analysis of our solutions, we begin by evaluating the entropy via the Iyer-Wald prescription \cite{Wald1993,IyerWald1994}. For a consistent application of this method, the scalar field $\psi$ must remain well-behaved at the event horizon $r_h$; otherwise, the procedure may lead to pathologies. To begin, we note that the action Lagrangian can be expressed in the form:
\begin{equation}
    \mathcal{L} = R + 4\alpha G^{ab}\nabla_a\psi\nabla_b\psi + (\text{Terms independent of }R^{ab}_{cd}).
\end{equation}

By employing the definition of the Einstein tensor $G^{ab}$, we can express the Lagrangian in the following form:
\begin{equation}
    \mathcal{L} = R(1 - 2\alpha\,(\nabla\psi)^2) + 4\alpha R^{ab}\nabla_a\psi\nabla_b\psi + (\text{Terms independent of }R^{ab}_{cd}).
\end{equation}

We define the tensor $P^{ab}_{cd}$ as:\begin{equation}16\pi P_{ab}^{cd} \equiv \frac{\partial \mathcal{L}}{\partial R_{cd}^{ab}},\end{equation}
where the factor of $16\pi$ is introduced to ensure that the entropy is expressed in its canonical form. Taking into account the antisymmetry of the Riemann tensor $R^{ab}_{cd}$ under the exchange of indices $a \leftrightarrow b$ and $c \leftrightarrow d$, the tensor $P^{ab}_{cd}$ can be obtained, after a straightforward calculation, as:\begin{equation}16\pi P_{ab}^{cd} = \left[ 1 - 2\alpha(\partial\psi)^2 \right] \delta_{[a}^{[c} \delta_{b]}^{d]} + 4\alpha\, \delta_{[a}^{[c} \nabla_{b]} \psi \nabla^{d]} \psi.\end{equation}

It is important to note that, in the present context of a three-dimensional spacetime, the Gauss-Bonnet term does not contribute to the tensor $P^{ab}_{cd}$ since the invariant $\mathcal{G}$ vanishes identically for $D=3$. Following the Iyer-Wald prescription, the black hole entropy is obtained by integrating this tensor over the bifurcation surface $\mathcal{H}$:\begin{equation}S = -2\pi \int_{\mathcal{H}} d^{D-2}x \sqrt{\gamma} \left[ P_{ab}^{cd} \hat{\epsilon}^{ab} \hat{\epsilon}_{cd} \right],
\end{equation}
where $\hat{\epsilon}_{ab}$ represents the binormal to the horizon, normalized such that $\hat{\epsilon}_{ab} \hat{\epsilon}^{ab} = -2$. For a static, spherically symmetric metric, the binormal resides strictly in the $(t, r)$ plane and is given by $\hat{\epsilon}_{ab} = 2t_{[a}r_{b]}$, where $t_a$ and $r_a$ are the components of the one-forms $dt$ and $dr$.The evaluation of the integrand involves the contraction of $P_{ab}^{cd}$ with the binormals, a step that reveals the "filtering" nature of the Wald formalism. While the first term in $16\pi P_{ab}^{cd}$ is proportional to the full kinetic term $(\partial\psi)^2$ in all directions, the second term is contracted directly with $\hat{\epsilon}^{ab} \hat{\epsilon}_{cd}$. Since the binormal is non-vanishing only in the $(t, r)$ subspace, this contraction isolates only the temporal and radial components of the scalar gradient. Consequently, these components cancel out precisely with the corresponding parts of the first term, leaving only the angular derivative as a non-trivial contribution. A simple computation then shows that:\begin{equation}P_{ab}^{cd} \hat{\epsilon}^{ab} \hat{\epsilon}_{cd} = -\frac{1}{8\pi} \left[ 1 - \frac{2\alpha}{r^2} (\partial_{\varphi} \psi)^2  \right] = -\frac{1}{8\pi}, \,\, \text{when} \,\, \psi = \psi(r).\end{equation}

In this case, we finally obtain the well-known formula for the entropy
\begin{equation}
    S = \frac{1}{4}\int_{\mathcal{H}} d^{D-2}x \sqrt{\gamma} = \frac{\mathcal{A}}{4},
\end{equation}
where the horizon ``area" is given by
\begin{equation}
    \mathcal{A} = \int_{\mathcal{H}} d^{D-2}x \sqrt{\gamma} .
\end{equation}
For $D=3$, this corresponds to the one-dimensional perimeter of a circle. It is important to note that the expression for the entropy as a function of the horizon radius $r_h$ will differ between the regular black hole and the black bounce cases. This distinction arises because the angular line elements for these two scenarios are different, directly affecting the integration over the bifurcation surface. Consequently, while the general form of the Wald entropy remains consistent, its specific dependence on the horizon parameters must be carefully evaluated according to the underlying geometry of the solution.

\subsection{Thermodynamics for the RBH case}
For the RBH case, the horizon area is given by
\begin{equation}
    \mathcal{A} = \int_{\mathcal{H}}d^{D-2}x \sqrt{\gamma} = \int_{0}^{2\pi}d\varphi\, r_h = 2\pi r_h \to S = \frac{\pi r_h}{2}.
\end{equation}
The Hawking temperature can be easily calculated following the procedures in Ref. \cite{Alencar:2026yyz}, which yields
\begin{equation}
    T =\frac{1}{2\pi K}\left(\frac{r_h}{\ell^2} - g'(r_h)\right).
\end{equation}
Explicitly
\begin{equation}
T=\frac{r_h}{2\pi K}\left(\frac{1}{\ell^2}-2\rho_0+\frac{2q^2}{(r_h^2+a^2)^2}\right).
\end{equation}
Note that the vacuum temperature is recovered when the sources vanish, namely for $q=\rho_0=0$.

In addition to the temperature, we can also calculate the heat capacity and the Helmholtz free energy, which are associated with the local and global thermodynamic stability, respectively. These quantities are defined as:\begin{equation}C = T \left( \frac{\partial S}{\partial T} \right)= T \frac{dS/dr_h}{dT/dr_h}, \quad \text{and} \quad F = M_{eff} - TS.\end{equation}

After a straightforward calculation, these expressions lead to
\begin{eqnarray}
    C(r_h) = \frac{\pi}{2}\left(\frac{r_h - \ell^2\, g'(r_h)}{1 - \ell^2\, g''(r_h)}\right), \quad \text{and} \quad F(r_h) = \frac{1}{4K}\left(\frac{3r_h^2}{\ell^2} + r_h\,g'(r_h) - 4g(r_h)\right).
\end{eqnarray}
Explicitly
\begin{eqnarray}
C(r_h)
&=&
\frac{\pi}{2}
\left[
\frac{
r_h
-\ell^2
\left(
2\rho_0 r_h
-\dfrac{2q^2 r_h}{(r_h^2+a^2)^2}
\right)
}{
1
-\ell^2
\left(
2\rho_0
-\dfrac{2q^2(a^2-3r_h^2)}{(r_h^2+a^2)^3}
\right)
}
\right],
\nonumber\\[0.4cm]
F(r_h)
&=&
\frac{1}{4K}
\left[ \left(\frac{3}{\ell^2 }- 2\rho_0\right)r_h^2
-\frac{2q^2(3r_h^2+2a^2)}{(r_h^2+a^2)^2}
\right].
\end{eqnarray}
Unlike the source-free case, where the black hole evaporates completely, the introduction of a source satisfying the symmetry $T^t_t = T^r_r$ modifies this behavior, allowing, in principle, the existence of a cold remnant ($T = 0$), whose radius is given by

\begin{equation}
    r_h^{cri} = \ell^2 g'(r_h^{cri}) \rightarrow 
r_h^{\mathrm{cri}}
=
\sqrt{
\sqrt{
\frac{2q^2\ell^2}{2\rho_0\ell^2-1}
}
-a^2
}.
\end{equation}

Moreover, by analyzing the expression for the heat capacity, it is possible to see that, in general, a phase transition may occur in this system when $2\rho_0\ell^2>1$. This transition is determined by the point at which the derivative of the temperature vanishes ($dT/dr_h = 0$), and therefore the heat capacity diverges, i.e., $C \to \infty$ when $\ell^2 g''(r_h) \to 1$. This critical point, denoted by $r_0$, is implicitly determined by

\begin{equation}
(2\rho_0\ell^2-1)(r_0^2+a^2)^3
=
2q^2\ell^2(a^2-3r_0^2).
\end{equation}

Unlike the extremal configuration, which is characterized by $T=0$ at $r_h^{\mathrm{cri}}$, the divergence of the heat capacity occurs at a distinct radius $r_0$, defined by $1-\ell^2 g''(r_0)=0$. By comparing both conditions, one finds that
\begin{equation}
r_0<r_h^{\mathrm{cri}}.
\end{equation}

Therefore, although the heat capacity formally diverges at $r_0$, this critical point is located beyond the extremal remnant radius and does not belong to the physical evaporating branch of the solution. Consequently, the black hole reaches the cold remnant configuration before any thermodynamic phase transition can actually take place.

The thermodynamic stability of the system can be investigated through the sign of the free energy. In particular, a Hawking--Page phase transition occurs whenever the free energy changes sign, namely when
\begin{equation}
F(r_h)=0.
\end{equation}

Explicitly, this condition yields
\begin{equation}
\left(\frac{3}{\ell^2}-2\rho_0\right)
r_h^2(r_h^2+a^2)^2
=
2q^2(3r_h^2+2a^2).
\end{equation}

To analyze the existence of such transition, let us consider the asymptotic behavior of the free energy. Near the point $r_h \to 0$, one finds
\begin{equation}
F(r_h\rightarrow0)
=
-\frac{q^2}{Ka^2}<0,
\end{equation}
showing that small black holes are thermodynamically favored.

On the other hand, for large values of the horizon radius,
\begin{equation}
F(r_h\rightarrow\infty)
\sim
\frac{1}{4K}
\left(
\frac{3}{\ell^2}-2\rho_0
\right)r_h^2.
\end{equation}

Therefore, the asymptotic sign of the free energy is completely determined by the combination
\begin{equation}
\frac{3}{\ell^2}-2\rho_0.
\end{equation}

If
\begin{equation}
2\rho_0\ell^2<3,
\end{equation}
the free energy becomes positive for sufficiently large horizon radius. Since it is negative near the origin, continuity implies the existence of at least one positive root satisfying
\begin{equation}
F(r_{HP})=0.
\end{equation}

Consequently, the system undergoes a Hawking--Page phase transition at the critical radius \(r_{HP}\).

Conversely, if
\begin{equation}
2\rho_0\ell^2>3,
\end{equation}
the free energy remains negative for all values of \(r_h\), implying that no Hawking--Page transition occurs.

Interestingly, the existence of the extremal remnant requires
\begin{equation}
2\rho_0\ell^2>1.
\end{equation}

Hence, both the cold remnant configuration and the Hawking--Page transition coexist in the parameter interval
\begin{equation}
1<2\rho_0\ell^2<3.
\end{equation}

In this regime, the matter sector not only regularizes the geometry and allows the formation of a zero-temperature remnant, but also introduces a rich thermodynamic structure characterized by nontrivial phase transitions.

\subsection{Thermodynamics for the BB case}

For the BB case, the horizon area for $D=3$ is given by
\begin{equation}
 \label{Sbb}   \mathcal{A} = \int_{\mathcal{H}}d^{D-2}x \sqrt{\gamma} = \int_{0}^{2\pi}d\varphi\, \Sigma(r_h) = 2\pi \Sigma(r_h) \to S = \frac{\pi \sqrt{r_h^2 + a^2}}{2}.
\end{equation}

The event horizon radius $r_h$ is determined by the condition
\begin{equation}
f(r_h)=0,
\end{equation}
where the metric function is \eqref{fbb}.

Imposing the horizon condition yields
\begin{equation}
C_1=\frac{\Sigma_h^2}{\ell^2},
\end{equation}
where
\begin{equation}
\Sigma_h=\sqrt{r_h^2+a^2}.
\end{equation}

Following the same asymptotic normalization adopted for the regular black hole solution, we define
\begin{equation}
K_{BB}=\sqrt{1+\frac{4\alpha}{\ell^2}},
\end{equation}
and introduce the effective mass parameter through the thermodynamic normalization discussed in Ref. \cite{Alencar:2026yyz}; thus, the integration constant is written as
\begin{equation}
C_1=K_{BB}M_{\rm eff}.
\end{equation}
Consequently,
\begin{equation}
M_{\rm eff}
=
\frac{\Sigma_h^2}{K_{BB}\ell^2}
=
\frac{r_h^2+a^2}
{K_{BB}\ell^2}.
\end{equation}
This prescription ensures that the thermodynamic quantities retain explicit dependence on the Gauss–Bonnet coupling and smoothly recover the previously obtained EGB-BTZ results in the limit $a\to 0$.

The Hawking temperature is obtained from the surface gravity,
\begin{equation}
T_{BB}
=
\frac{f'(r_h)}
{4\pi K_{BB}}.
\end{equation}

Using $\Sigma'(r)=r/\Sigma$, a direct differentiation of the metric function gives
\begin{equation}
f'(r_h)=\frac{2r_h}{\ell^2}.
\end{equation}
Therefore,
\begin{equation}
T_{BB}
=
\frac{r_h}
{2\pi K_{BB}\ell^2}.
\end{equation}

The heat capacity is defined as
\begin{equation}\label{heat-capacity-BB}
C_{BB}
=
T_{BB}
\frac{\partial S_{BB}/\partial r_h}
{\partial T_{BB}/\partial r_h}.
\end{equation}
Using Eq.\eqref{Sbb}
one finds
\begin{equation}
\frac{\partial S_{BB}}
{\partial r_h}
=
\frac{\pi r_h}
{2\sqrt{r_h^2+a^2}},
\end{equation}
while
\begin{equation}
\frac{\partial T_{BB}}
{\partial r_h}
=
\frac{1}
{2\pi K_{BB}\ell^2}.
\end{equation}
Substituting these expressions into Eq. \ref{heat-capacity-BB} yields
\begin{equation}
C_{BB}
=
\frac{\pi r_h^2}
{2\sqrt{r_h^2+a^2}}.
\end{equation}

The Helmholtz free energy is given by
\begin{equation}
F_{BB}
=
M_{\rm eff}
-
T_{BB}S_{BB}.
\end{equation}
Substituting the previous results, we obtain
\begin{equation}
F_{BB}
=
\frac{1}
{K_{BB}\ell^2}
\left[
r_h^2+a^2
-
\frac{r_h\sqrt{r_h^2+a^2}}{4}
\right].
\end{equation}

In the limit $a\rightarrow0$, one recovers the corresponding BTZ-EGB expressions,
\begin{equation}
T_{BB}
\rightarrow
\frac{r_h}
{2\pi K_{BB}\ell^2},
\end{equation}
\begin{equation}
C_{BB}
\rightarrow
\frac{\pi r_h}{2},
\end{equation}
and
\begin{equation}
F_{BB}
\rightarrow
\frac{3r_h^2}
{4K_{BB}\ell^2}.
\end{equation}

\section{Conclusions}

In this work, we investigated the construction of nonsingular compact objects within the regularized lower-dimensional Einstein--Gauss--Bonnet framework. Starting from the BTZ-like vacuum solution previously obtained in this theory, we emphasized that the inclusion of Gauss--Bonnet corrections gives rise to a genuine curvature singularity at the origin, despite the fact that the original BTZ geometry is free from such pathologies. This feature provides an interesting example in which higher-curvature corrections worsen, rather than improve, the ultraviolet behavior of the spacetime.

To overcome this difficulty, we reconstructed matter sectors capable of restoring regularity while preserving the BTZ-like asymptotic structure. We first showed that nonlinear electrodynamics offers a natural mechanism for generating regular black-hole configurations. By deriving the corresponding matter sources directly from the gravitational field equations, we obtained explicit solutions whose curvature invariants remain finite throughout the entire spacetime. The same reconstruction strategy was subsequently extended to Simpson--Visser black-bounce geometries, demonstrating that smooth throat configurations can also be consistently embedded within the lower-dimensional Einstein--Gauss--Bonnet framework.

A central result of this work is that singularity resolution can also be achieved without resorting to nonlinear electromagnetic interactions. By formulating a unimodular extension of lower-dimensional Einstein--Gauss--Bonnet gravity, we showed that standard Maxwell fields are capable of supporting nonsingular geometries through a dynamical exchange between the vacuum and matter sectors. In this scenario, the spacetime-dependent cosmological function acts as an effective energy reservoir, generating the currents required to sustain the geometry and providing an alternative mechanism for singularity avoidance. This construction extends recent results obtained in four-dimensional unimodular gravity to the lower-dimensional Einstein--Gauss--Bonnet context.

The thermodynamic analysis revealed that the presence of matter sources substantially modifies the evaporation process. For the regular black-hole solutions, the interplay between the vacuum contribution and the nonlinear electromagnetic sector leads to a rich thermodynamic structure, including the possibility of cold remnants and Hawking--Page phase transitions. For the black-bounce configurations, we derived the corresponding thermodynamic quantities and showed that the bounce parameter deforms the entropy, heat capacity, and free energy while smoothly recovering, in the appropriate limit, the Einstein--Gauss--Bonnet BTZ thermodynamics obtained within the normalized framework introduced in our previous analysis, where the physical quantities explicitly depend on the Gauss--Bonnet coupling through the factor $K$ \cite{Alencar:2026yyz}. These results illustrate how different regularization mechanisms can leave distinct imprints on the thermodynamic properties of lower-dimensional compact objects.

Overall, our findings indicate that lower-dimensional Einstein--Gauss--Bonnet gravity provides a valuable laboratory for investigating the interplay between higher-curvature corrections, nonsingular geometries, modified conservation laws, and black-hole thermodynamics. Beyond furnishing explicit examples of regular black holes and black bounces, the unimodular construction developed here suggests that singularity resolution may emerge from dynamical vacuum--matter interactions rather than exclusively from nonlinear matter sectors. Future investigations may explore rotating extensions of these solutions, quasinormal-mode spectra, holographic aspects, and possible connections with quantum-gravity-inspired regularization mechanisms.

\acknowledgments{We acknowledge the financial support provided by the Conselho Nacional de Desenvolvimento Científicoe Tecnológico (CNPq), Fundação Cearense de Apoio ao Desenvolvimento Científico e
Tecnológico (FUNCAP) and Coordena\c c\~{a}o de Aperfei\c coamento de Pessoal de N\'{i}vel Superior - Brasil (CAPES) - Finance Code 001.
}

\subsection*{Appendix A: Explicit Source Functions for the Black Bounce Case}\label{apendice}

The potential associated with the scalar field $\phi$ (Eq. \eqref{Vr}) is explicitly given as a function of $r$ as follows:
\begin{equation}
\begin{aligned}
V(r) = &-\frac{a^2}{1920 C_1^2 \ell^6 (a^2 + r^2)^3 \alpha^3} \Bigg\{ 16 \ell^2 (a^2 + r^2) \sqrt{\frac{4 r^2 \alpha + a^2 (\ell^2 + 4 \alpha) + \ell^2 (r^2 - 4 C_1 \alpha)}{\ell^2 (a^2 + r^2)}} \\
&\times \Big[ 3 a^6 (\ell^2 + 4 \alpha)^2 + 2 a^4 (\ell^2 + 4 \alpha) (12 r^2 \alpha + \ell^2 (3 r^2 - 2 C_1 \alpha)) \\
&+ 10 C_1 \ell^2 r^2 \alpha (-4 r^2 \alpha + \ell^2 (-r^2 + C_1 \alpha)) \\
&+ a^2 (48 r^4 \alpha^2 + 8 \ell^2 r^2 \alpha (3 r^2 - 7 C_1 \alpha) + \ell^4 (3 r^4 - 14 C_1 r^2 \alpha + 28 C_1^2 \alpha^2)) \Big] \\
&- 15 \Big[ a^8 (\ell^2 + 4 \alpha)^3 + a^6 (\ell^2 + 4 \alpha)^2 (12 r^2 \alpha + \ell^2 (3 r^2 - 10 C_1 \alpha)) \\
&+ 2 C_1 \ell^2 r^2 \alpha (-80 r^4 \alpha^2 - 8 \ell^2 r^2 \alpha (5 r^2 - 8 C_1 \alpha) + \ell^4 (-5 r^4 + 16 C_1 r^2 \alpha + 16 C_1^2 \alpha^2)) \\
&+ a^4 (\ell^2 + 4 \alpha) (48 r^4 \alpha^2 + 24 \ell^2 r^2 \alpha (r^2 - 5 C_1 \alpha) + \ell^4 (3 r^4 - 30 C_1 r^2 \alpha + 32 C_1^2 \alpha^2)) \\
&+ a^2 (64 r^6 \alpha^3 + 48 \ell^2 r^4 \alpha^2 (r^2 - 10 C_1 \alpha) + 4 \ell^4 r^2 \alpha (3 r^4 - 60 C_1 r^2 \alpha + 64 C_1^2 \alpha^2) \\
&+ \ell^6 (r^6 - 30 C_1 r^4 \alpha + 64 C_1^2 r^2 \alpha^2 - 32 C_1^3 \alpha^3)) \Big] \Bigg\}.
\end{aligned}
\end{equation}
 and as a function of $\phi$ (Eq. \eqref{vp}):
\begin{equation}
\begin{aligned}
V(\phi) = &\frac{a^2 (\ell^2 + 4 \alpha)^2 \left[ a^2 (\ell^2 + 4 \alpha) - 10 C_1 \ell^2 \alpha \right]}{128 C_1^2 \ell^6 \alpha^3} + \left( \frac{1}{\ell^2} + \frac{1}{4 \alpha} \right) \cos^2\phi - \frac{C_1 \cos^4\phi \cos(2\phi)}{4 a^2} \\
&- \frac{1}{480 C_1^2 \ell^4 \alpha^3} \sqrt{1 + \frac{4 \alpha}{\ell^2} - \frac{4 C_1 \alpha \cos^2\phi}{a^2}} \\
&\times \Big\{ 12 a^4 (\ell^2 + 4 \alpha)^2 - 28 a^2 C_1 \ell^2 \alpha (\ell^2 + 4 \alpha) + 47 C_1^2 \ell^4 \alpha^2 \\
&+ C_1 \ell^2 \alpha \left[ 4 \left( 3 a^2 (\ell^2 + 4 \alpha) + 14 C_1 \ell^2 \alpha \right) \cos(2\phi) + 9 C_1 \ell^2 \alpha \cos(4\phi) \right] \Big\}.
\end{aligned}
\end{equation}

The functions $\mathcal{L}_F$
 and $\mathcal{L}$ of the NED source for the black bounce case studied in Section \ref{nedbb} are explicitly given as functions of $r$ by
\begin{equation}
\begin{split}
    \mathcal{L}_F(r) &= \frac{32 q^2}{(a^2+r^2)} \Biggl\{ 
    -\frac{8 C_1 r^2 \left[r^2+r (a^2+r^2)^{3/2}-(a^2+r^2)^2\right] (1-\sqrt{\Delta})}{(a^2+r^2)^3 \sqrt{\Delta}} \\
    &\quad + \frac{4 (\sqrt{\Delta}-1)}{\alpha} - \frac{4 r^2 (\sqrt{\Delta}-1)}{(a^2+r^2) \alpha} + \frac{2 a^4 (\sqrt{\Delta}-1)^2}{(a^2+r^2) \alpha} \\
    &\quad + \frac{4 r^2 \left[r^2+r (a^2+r^2)^{3/2}-(a^2+r^2)^2\right] (\sqrt{\Delta}-1)^2}{(a^2+r^2)^2 \alpha} \\
    &\quad + \frac{(\sqrt{\Delta}-1)^2}{(a^2+r^2)^{5/2} \alpha} \Bigl[ -2 r^4 \sqrt{a^2+r^2} + a^6 \left(3 r-2 \sqrt{a^2+r^2}\right) \\
    &\qquad\qquad + 2 a^4 r^2 \left(3 r-2 \sqrt{a^2+r^2}\right) \\
    &\qquad\qquad + a^2 \left(3 r^5+4 r^2 \sqrt{a^2+r^2}-2 r^4 \sqrt{a^2+r^2}\right) \Bigr] \\
    &\quad + 4 \left[ \frac{8 C_1^2 r^2 \alpha}{(a^2+r^2)^3 \Delta^{3/2}} - \frac{2 C_1}{(a^2+r^2) \sqrt{\Delta}} + \frac{1-\sqrt{\Delta}}{\alpha} \right] \Biggr\}^{-1},
\end{split}
\end{equation}

\begin{equation}
\begin{split}
    \mathcal{L}(r) &= -\frac{1}{\ell^2} - \frac{4 C_1^2 r^2 \alpha}{(a^2+r^2)^3 \Delta^{3/2}} + \frac{2 C_1}{(a^2+r^2) \sqrt{\Delta}} + \frac{\sqrt{\Delta}-1}{\alpha} \\
    &\quad + \frac{C_1 r^2 \left(\sqrt{a^2+r^2}-r\right) (\sqrt{\Delta}-1)}{(a^2+r^2)^{3/2} \sqrt{\Delta}} - \frac{a^4 (\sqrt{\Delta}-1)^2}{4 (a^2+r^2) \alpha} \\
    &\quad - \frac{a^2 r (\sqrt{\Delta}-1)^2}{8 \sqrt{a^2+r^2} \alpha} + \frac{r^2 \left(1 - \frac{r}{\sqrt{a^2+r^2}}\right) (\sqrt{\Delta}-1)^2}{4 \alpha} \\
    &\quad + \frac{(a^2+r^2) \left(1 - \frac{r}{\sqrt{a^2+r^2}}\right) (\sqrt{\Delta}-1)^2}{4 \alpha} \\
    &\quad + \frac{a^2}{2(a^2+r^2)^2} \left[ -\frac{2 C_1}{\sqrt{\Delta}} - \frac{(a^2+r^2)(\sqrt{\Delta}-1)}{\alpha} \right].
\end{split}
\end{equation}

\bibliography{ref}
\end{document}